\documentclass[twocolumn,showpacs,preprintnumbers,amsmath,amssymb]{revtex4-1}

\usepackage{graphicx}
\usepackage{dcolumn}
\usepackage{bm}
\usepackage{color}
\usepackage[latin1]{inputenc}

\usepackage{ulem}  

\begin{document}

\preprint{APS/123-QED}

\title{Degassing cascades in a shear-thinning viscoelastic fluid}

\author{Val\'erie Vidal, Fran\c{c}ois Soubiran, Thibaut Divoux and Jean-Christophe G\'eminard}
\affiliation{Universit\'e de Lyon, Laboratoire de Physique, \'Ecole Normale Sup\'erieure de
Lyon, CNRS, 46 All\'ee d'Italie, 69364 Lyon cedex 07, France.}

\date{\today}

\begin{abstract}
We report the experimental study of the degassing dynamics through a thin layer of 
shear-thinning viscoelastic fluid (CTAB/NaSal solution), when a constant air flow is imposed at its bottom. 
Over a large range of parameters, the air is periodically released through series of successive bubbles, 
hereafter named {\it cascades}. Each cascade is followed by a continuous degassing, lasting for several seconds, 
corresponding to an open channel crossing the fluid layer. The periodicity between two cascades 
does not depend on the injected flow-rate. Inside one cascade, the properties of the overpressure signal
associated with the successive bubbles vary continuously. The pressure threshold above which the fluid
starts flowing, fluid deformation and pressure drop due to degassing through the thin fluid layer can be 
simply described by a Maxwell model. We point out that monitoring the evolution 
inside the cascades provides a direct access to the characteristic relaxation time associated with the 
fluid rheology.
\end{abstract}

\maketitle

\section{Introduction}

Non-Newtonian fluids peculiar dynamics has focused a large attention over the last decades. 
Their ability to flow or break \cite{Gladden07,Tabuteau09}, to exhibit viscous or elastic behavior \cite{Bird87},
thixotropy effects \cite{Mewis79,Barnes97,Mewis09}
or to sustain loads without flowing \cite{Coussot02, Moller06} opened a wide range of applications,
including cosmetics, food industry, environment and biology. Among them, micellar fluids differ
from polymers by their internal dynamics: they continuously break and restructurate \cite{Shikata88,Cates90b,Gelbart94}, 
and their internal kinetics determines their length distribution - and, hence, their macroscopic properties \cite{Cates90a,Berret97}.

The behavior of bubbles rising through such fluids is complex, due, in particular, to the non-trivial coupling 
between the bubble and the fluid rheology \cite{Chhabra07}. This has prevented, up to now, a complete theoretical 
description of the system, and favored experimental studies. Due to the viscoelastic properties, 
the bubble shape is generally elongated \cite{Bird87,Belmonte00,Handzy04,Chhabra07,Divoux08b}, 
its tail ends by a cusp \cite{Belmonte00,Chhabra07}, and both its geometry and  
velocity oscillate during its rising through the fluid  \cite{Belmonte00,Jayaraman03,Handzy04}.
In shear-thinning fluids, the local perturbation due to a rising bubble - or a falling sphere - 
creates a negative wake \cite{Hassager79} and a corridor of reduced viscosity \cite{Daugan02a,Daugan02b}. 
As a consequence, successive bubbles may interact one with the other \cite{Li01,Lin03,Li04b,Divoux09},
if the emission period is shorter than the time for the perturbation created by the bubble to vanish.

An extensive study of the degassing regimes when a constant air flux is imposed
at the bottom of a complex, yield-stress fluid column (gel or immersed granular material) has revealed
the existence of three different regimes \cite{Gostiaux02,Divoux09,Varas09}. 
On the one hand, at low flow-rate, bubbles rise independently one from the other. 
On the other hand, at high flow-rate, an open channel connects the bottom nozzle to 
the fluid surface; this channel develops instabilities, forming a `bubble chain' 
\cite{Kliakhandler02}. Finally, at intermediate flow-rate, the system spontaneously 
oscillates between the two previous regimes, exhibiting a complex intermittent
dynamics \cite{Divoux09}.

\begin{figure}[b]
\begin{center}
\includegraphics[width=0.9\columnwidth]{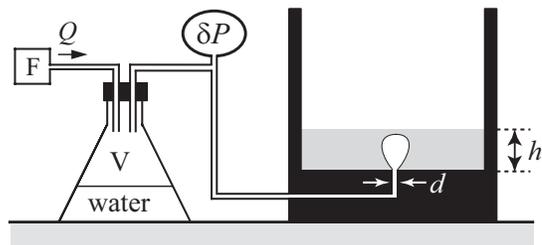}
\end{center}
\caption{{\bf Experimental setup.}
A constant air-flow is supplied at the base of a thin layer of micellar fluid by an
air-flow controller {\bf F} (flow-rate $Q$), via a chamber of volume $V$. $h$ is of the
order of a typical bubble diameter (a few millimeters to centimeters).}
\label{fig:expsetup}
\end{figure}

In this article, we extend these previous works to the case of a fluid without yield stress,
and for a {\it thin} fluid layer,
i.e., when the fluid height is smaller than the typical size of a bubble.
We report the existence of a peculiar degassing regime in a thin layer of micellar fluid (CTAB/NaSal mixture).
When a constant air flow is imposed at its bottom, in the bubbling
regime, the air is released through periodic series of successive bubbles, hereafter named {\it cascades}. 
When the layer is thin enough, the cascades merely consists in successive opening
and closing of the fluid layer above the air injection point. For convenience, these periodic
degassing apertures will also be named {\it bubbles} in the following.
We measure the overpressure at the base of the fluid column, at the injection point. The main characteristics
of the cascades (periodicity, maximum overpressure) are analyzed in regards with 
the fluid rheology. We show that inside a cascade, the bubble properties vary
continuously: the mean overpressure drops, whereas the emission time increases.
The transition between two regimes inside the cascades is directly linked to the fluid rheology
and, in particular, provides a direct measurement of the viscoelastic characteristic time associated with the 
Maxwell fluid.

\section{Experimental setup}

The experimental cell consists of a cylinder made of plexiglass (diameter 74~mm, 
height 270~mm), filled with the fluid up to a height $h$ (Fig.~\ref{fig:expsetup}). 
Air is injected at constant flow-rate $Q$ (from 0.17 to 1.72~mL/s)
through an injection hole (diameter $d=2$~mm) at the bottom of the fluid column.
The air-injection system consists of a mass-flow controller (Bronkhorst, Mass-Stream 
series  D-5111) connected to a chamber of volume $V$, from which the air flows 
at the column bottom. 
The volume of the chamber can be easily tuned by changing the water level into
the chamber (see Fig.~\ref{fig:expsetup}). The water itself makes it possible to 
inject humid air inside the fluid, thus avoiding any drying of the sample over the
experimental time.
A differential pressure sensor (223 BD-00010 AB, MKS Instruments) measures the 
variations of the overpressure $\delta P$ inside the chamber, corresponding to the 
pressure variations at the bottom of the cell.

The fluid is a semi-dilute micellar system obtained by a mixing at equimolar 
concentration, inside pure water, sodium salicylate (NaSal, Sigma Aldrich) 
and hexadecyltrimethylammonium bromide (CTAB, Sigma Aldrich).
The mixture of this two chemical components causes the formation of a network
of giant entangled micelles, which break down and reform continuously \cite{Shikata88}.
On a macroscopic point of view, the fluid exhibits shear-thinning, viscoelastic properties \cite{Shikata87,Inoue05}, 
which can be tuned by varying the fluid concentration $c$ (from 0.03 to 0.5~mol.L$^{-1}$).
The rheology of these well-controlled mixtures is characterized by rheometer measurements
(see Appendix~\ref{app:rheology} and \ref{app:modulii}).
The height $h$ of the thin layer ranges from 5 to 35~mm. Unless specified, the results
presented here are for a typical height of 5~mm and a concentration $c=0.1$~mol.L$^{-1}$.

\section{Degassing regimes}

When varying the system parameters ($c,V,Q,h$), we observe three different degassing 
regimes through the thin layer: a {\it bubbling} regime, for which bubbles are emitted one after the other;
an {\it open channel} regime, for which the system is able to sustain a stable channel connecting
the injection nozzle at the base of the fluid column to the fluid free surface; and 
an {\it intermittent} regime, for which the system alternates spontaneously
between the bubbling and the open channel regime -- this latter pinching off intermittently. 
Similar degassing regimes have been reported in yield-stress fluids: gels \cite{Divoux09} or in immersed 
granular media \cite{Gostiaux02,Varas09}. 

\begin{figure}[t]
\begin{center}
\includegraphics[width=0.8\columnwidth]{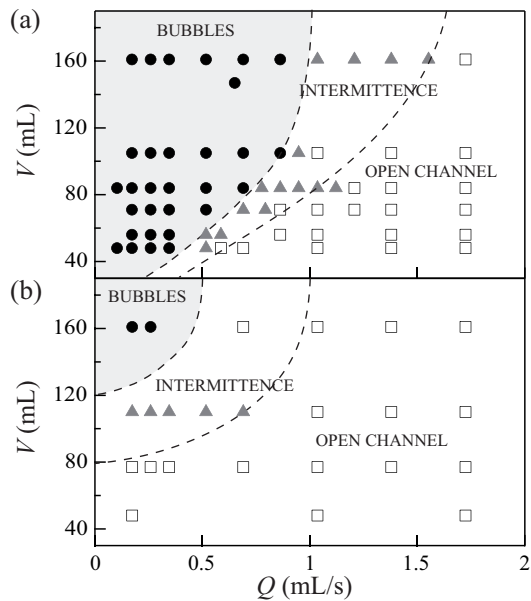}
\end{center}
\caption{ \label{fig:regimes}
Phase diagram of the different degassing regimes depending on the parameters ({\it V,Q}) for
$c=0.1$~mol.L$^{-1}$ (a) and $c=0.5$~mol.L$^{-1}$ (b)
[($\bullet$) bubbles, ($\blacktriangle$) intermittence, ($\square$) open channel].
Dashed lines are eyeleads. The gray zone correspond to the space of parameters
where bubble cascades are observed [$h=5$~mm].}
\end{figure}

Typically, when increasing the
injected air flow-rate $Q$, all other parameters being constant, the system goes
from the bubbling, intermittent and finally open channel regime (Fig.~\ref{fig:regimes}).
When increasing the fluid concentration, it becomes easier for the system to sustain
an open channel, and it is necessary to go to higher volumes $V$ and smaller flow rate $Q$
to observe the bubbling regime (Fig.~\ref{fig:regimes}a and b) \cite{RMQ}. On the contrary,
increasing the fluid column height makes it more difficult to open a channel, and shifts
the regime boundaries toward smaller volumes and higher flow rates. 
Note that the precise boundary between two different regimes is difficult to determine,
due to the finite acquisition time (the system may not have the time to switch from bubbles
to an open channel, or vice-versa).

The formation of an open-channel can be qualitatively explained. Indeed, due to the fluid 
shear-thinning properties (Appendix~\ref{app:rheology}), when a bubble rises through the fluid, 
its wake is characterized by a local viscosity smaller than the surrounding fluid. If the flow rate
is large enough, the following bubble will rise through a fluid with a smaller effective viscosity --
and thus, will rise faster, and so on, until the system is able to sustain an open channel
through the fluid column. Note that this channel does not resemble a cylinder, but rather
a bubble chain, similar to previous observations in yield-stress and non yield-stress fluids
\cite{Kliakhandler02,Divoux09}.

\begin{figure*}[t]
\begin{center}
\includegraphics[width=1.2\columnwidth]{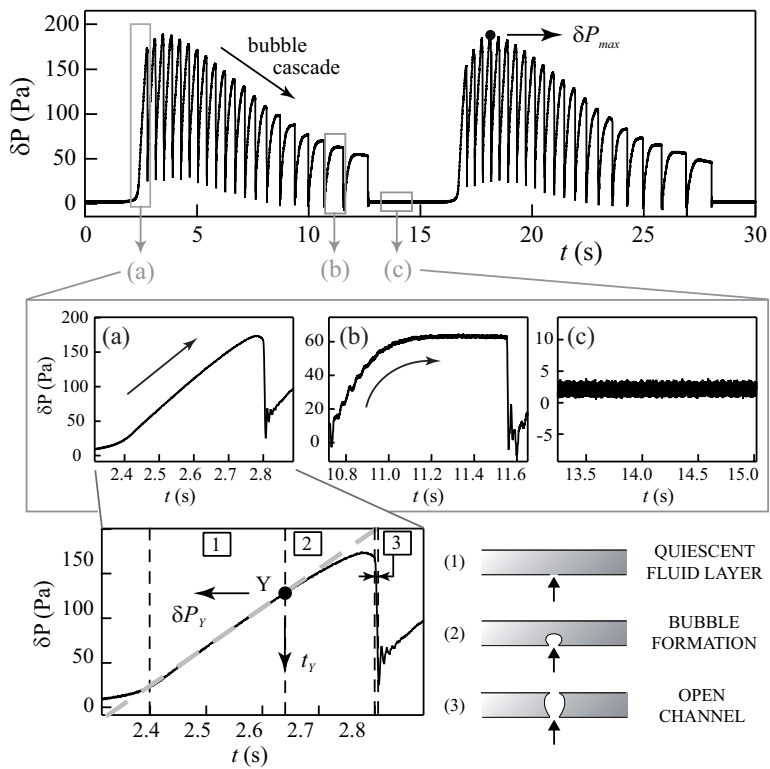}
\end{center}
\caption{ \label{fig:cascade}
{\it Top:} Pressure signal displaying the gas discharge via bubble cascades.
$\delta P_{max}$ indicates the maximum overpressure reached in the cascades.
{\it Middle:} (a) The first bubbles in the cascade exhibit a linear pressure increase.
The fluid starts flowing only when the bubble is about to be emitted (pressure jump).
(b) The last bubbles in the cascade exhibit a curved pressure increase, characteristic
of fluid deformation (flowing) and bubble growth. (c) Between two bubble cascades,
an open channel connects the air nozzle to the fluid free surface. The overpressure
is constant, almost equal to zero (see text). 
{\it Bottom:} Detail of the pressure signal for a bubble formation and emission inside
 a cascade: (1) linear pressure increase; the dashed gray line corresponds to the 
 linear pressure increase in the fixed volume $V$,  $\delta P=(P_0/V)Qt$, without any 
 ajustable parameters (see text); (2) bubble formation and growth; 
 (3) the bubble pierces the free surface and the gas is released. 
 Note the pressure oscillations subsequent to the bubble emission
[$c=0.1$~mol.L$^{-1}$, $V=147$~mL, $Q=0.65$~mL/s].}
\end{figure*}

\section{Bubble cascades}
\label{sec:cascades}

In a wide range of fluid concentration (from $c=0.04$ to $0.14$~mol/L),
when the fluid layer height $h$ is of the order of the size of a bubble (typically, $h<10$~mm), we do not
observe, in the bubbling regime, successive bubbles but rather periodic series of bubbles,
hereafter named {\it bubble cascades}. These cascades are clearly observed in the overpressure 
signal $\delta P$ recorded at the bottom of the fluid column (Fig.~\ref{fig:cascade}, top). 

The pressure signal corresponding to one cascade exhibits successive rises and drops, 
each of them corresponding to the pressure increasing in the chamber, followed by a bubble emission
(Fig.~\ref{fig:cascade} middle, a and b). After each bubble cascade, the overpressure
remains at $\delta P \sim 0$ for a few seconds, before the next bubble cascade (Fig.~\ref{fig:cascade} middle, c). 
During this time interval, the air escapes continuously through the thin fluid layer. 
The value of the overpressure is given by the charge loss of the air flow through the hole 
(roughly a few Pa, Fig.~\ref{fig:cascade}, middle, c). This hole then suddenly closes, and
the next cascade starts. The sequence consisting of a bubble cascade followed
by a hole opened through the fluid layer repeats periodically in time. The main goal of
this work is to describe thoroughly the cascades properties.

\begin{figure}[t] 
\centerline{\includegraphics[width=0.9\columnwidth]{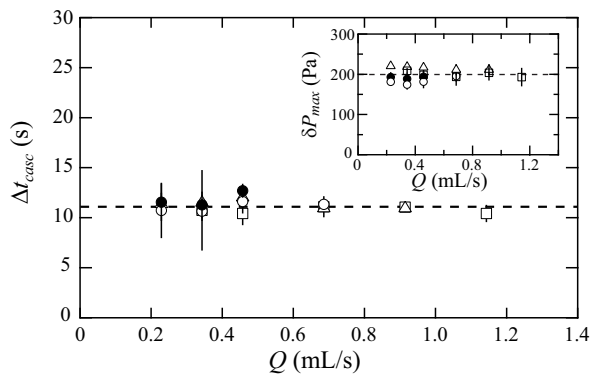}}
 \caption{Time interval $\Delta t_{casc}$ separating a sequence `bubble cascade + hole',
 as a function of the air flow-rate $Q$.
{\it Inset :} Maximum overpressure $\delta P_{max}$ reached in the cascades (see Fig.~\ref{fig:cascade}).
(Symbol, $V$ [mL]): ($\bullet$,~56), ($\circ$,~71), ($\triangle$, ~83), ($\square$,~161).
Dashed lines indicate the mean value of all the data.}
 \label{fig:DtPmax} 
 \end{figure}
 
\begin{figure}[t] 
\centerline{\includegraphics[width=0.9\columnwidth]{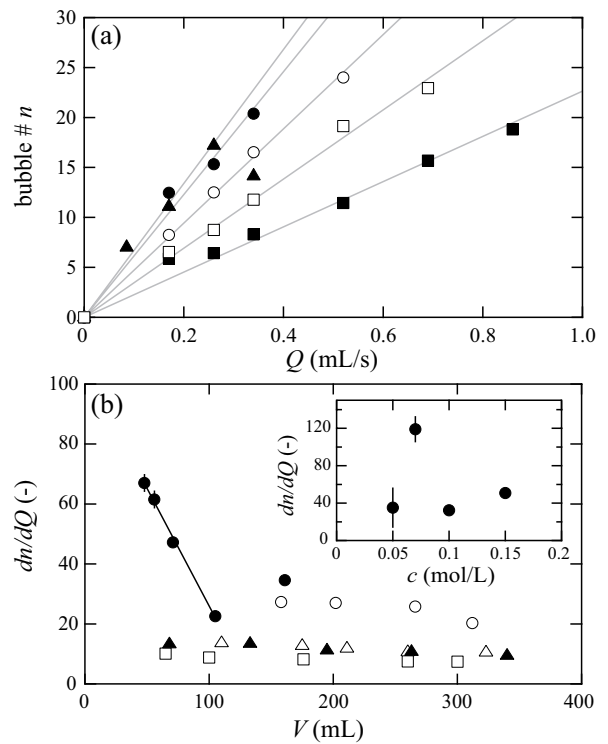}}
 \caption{
(a) Number of bubbles per cascade $n$ as a function of the air flow rate $Q$.
(Symbol, $V$ [mL]): ($\blacktriangle$,~48), ($\bullet$,~56), ($\circ$,~71), ($\blacksquare$, ~105), ($\square$,~161).
The gray lines correspond to the linear interpolation for each series of experiments ($V$ fixed)
[$h=5$~mm].
(b) $dn/dQ$ as a function of $V$. The series reported in (a) display a linear, decreasing 
relationship up to $V=105$~mL. For larger fluid layer height, the slope is almost constant. 
(Symbol, $h$ [mm]): ($\bullet$,5); ($\circ$,8); ($\triangle$,18); ($\square$,27); ($\blacktriangle$,35).
{\it Inset:} $dn/dQ$ as a function of the fluid concentration [$V=147$~mL, $Q=0.65$~mL/s, $h=5$~mm]. 
No apparent relationship is found.}
 \label{fig:bubnum} 
 \end{figure}

 The general properties of the cascades are reported in Figure~\ref{fig:DtPmax}. 
 On the one hand, we observe
 that the time interval between each sequence (cascade + hole) is very stable
 and, over the range of parameters explored, does not depend neither on the injected 
 flow rate $Q$, nor on the chamber volume $V$.
 The same observation is reported for the maximum overpressure reached inside each
 cascade (Fig.~\ref{fig:DtPmax}, inset).
On the other hand, the number of bubbles emitted per cascade, $n$, depends linearly
on the injected flow rate (Fig.~\ref{fig:bubnum}a). The associated slope,
$dn/dQ$, varies with the chamber volume $V$. 
For $h=5$~mm and small volumes, we report a linear, decreasing relationship 
$dn/dQ = -\zeta V$, where $\zeta$ is a constant (Fig.~\ref{fig:bubnum}b, black dots). 
For large values of $V$ ($>105$~mL) or larger height, however, $dn/dQ$ is independent 
of the chamber volume. These results can be interpreted as follows.

For small $h$, the layer is of the order
of a bubble height, and whenever a bubble is emitted, it pierces the layer. Via the open channel 
thus formed, the overpressurized air escapes the chamber. For small chamber volume $V$, 
the chamber reservoir quickly empties, and a dependence on $V$ is clearly seen. When
$V$ reaches higher values, it acts as a pressure reservoir and $dn/dQ$ remains roughly constant.

For larger $h$, the air is no longer able to pierce the fluid layer, and bubbles are emitted in the fluid, 
rise and burst without connecting anymore the injection nozzle to the fluid free surface. In this
case, the overpressurized air trapped in the chamber cannot escape directly through the fluid, 
but via the successive bubbles. The bubble size is fixed by the nozzle size and the fluid rheological
properties, and $dn/dQ$ is independent of $V$. 

The total volume of gas emitted during a cascade can be written $V_T \sim Q \Delta t_{casc}$.
As the number of bubbles emitted per cascade always depends linearly on $Q$, we 
can write $n=\alpha Q$, where $\alpha=dn/dQ$ is a constant for a given series of 
experiment ($c$, $h$ and $V$ fixed). We thus find that the average gas volume emitted
per bubble, $\langle v_b \rangle=V_T/n$, is constant. For $h>5$~mm, this constant does
not depend anymore on $V$ (Fig.~\ref{fig:bubnum}b).

Inside a cascade, however, the bubble properties (maximum overpressure and emission duration) 
vary continuously (Fig.~\ref{fig:cascade}, top and middle). In the next two sections, we 
investigate these variations, and see which informations they bring on the system.

\section{Evolution inside a cascade}

During the release of a single bubble, the overpressure $\delta P$ exhibits 
three different stages (Fig.~\ref{fig:cascade}, bottom). 
First, we observe a linear pressure increase (Fig.~\ref{fig:cascade}, bottom, region~1). 
The overpressure signal then departs from the linear tendency  (Fig.~\ref{fig:cascade}, bottom, region~2)
until the bubble is emitted  (sudden pressure drop, Fig.~\ref{fig:cascade}, bottom, region~3).
In this section, we describe each part of the pressure signal, and show that a simple Maxwell 
model can account for the different observations.

\subsection{Linear pressure increase}

When submitted to a sudden stress (pressure increase), at a short time scale, the 
CTAB/NaSal mixture does not flow, 
and the system is equivalent to a chamber of volume $V$ continuously filled by a gas flow $Q$. 
The overpressure is given by 
\begin{equation}
\delta P= \left( \frac{P_0}{V} \right) Qt
\end{equation} 
where $P_0=10^5$~Pa denotes the 
atmospheric pressure. The experimental slope is consistent with this linear pressure increase 
(dashed gray line, Fig.~\ref{fig:cascade} bottom, region~1), without any adjustable parameters.
In the following section, we estimate the threshold pressure $\delta P_Y$ above which the fluid 
starts flowing when the pressure at its bottom increases.

\begin{figure}[t]
\begin{center}
\includegraphics[width=0.8\columnwidth]{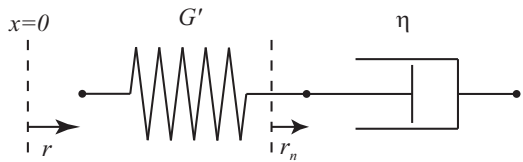}
\end{center}
\caption{Maxwell model for the CTAB/NaSal. This simple model consists in an ideal elastic spring
(elastic modulus $G'$) and a dashpot representing the viscous loss (viscosity $\eta$).
$r$ and $r_n$ indicate the displacements generated by the rising bubble and the dashpot shift after
the $n^{th}$ bubble, respectively.}
\label{fig:Maxwell}
\end{figure}

\subsection{Threshold pressure $\delta P_Y$}
\label{sec:dPY}

After a certain time $t_Y$, associated with an overpressure $\delta P_Y$ 
(point $Y$, Fig.~\ref{fig:cascade}, bottom), the overpressure departs from its linear increase. 
At this point, the fluid starts flowing, and a bubble is nucleated and grows at the tip of the injection 
hole (Fig.~\ref{fig:cascade}, bottom, region~2). In order to estimate this threshold pressure,
we describe the fluid with a Maxwell model, consisting of an ideal elastic spring 
attached to an ideal dashpot (Fig.~\ref{fig:Maxwell}). This simple model represents the viscoelastic behavior of energy 
storage and viscous loss, which can be quantified by the elastic ($G'$) and viscous ($G"$) modulus,
respectively. These modulii can be considered constant as a function of the applied stress
in a given frequency range ($\omega \sim 1$~Hz) representative of the frequency of 
bubbles rising through the fluid. The modulii are estimated from the plateau obtained from oscillation 
measurements (see Appendix~\ref{app:modulii}, Fig.~\ref{fig:moduliiSigma}) to 
$G' \sim 50$~Pa and $G'' \sim 10$~Pa.

The equation describing the opening of a hole of radius $r$ due to a bubble rising in the fluid layer can
be written:
\begin{equation}  \label{eq:motion}
\xi \rho r \frac{d^2r}{dt^2} = - \frac{2\gamma}{r} - \alpha G' (r-r_n) + \delta P - \rho g \Delta h
\end{equation} 
where $\xi$ is a constant.
The first term in the right-hand side represents the closing force due to the fluid surface tension $\gamma$, where 
$\gamma \simeq 40$~mN/m. The second term describes the elastic force
which tends to shift the spring back to its initial length, where $\alpha$ is a constant which can be approximated 
to the inverse of the nozzle radius $2/d$. $\delta P$ is the pressure inside the chamber (Fig.~\ref{fig:expsetup}) 
-- and, thus, inside the bubble which starts being generated at the injection nozzle. Finally,
the last term quantifies the weight associated with the fluid layer height $\Delta h$ above the newborn bubble.
At the limit where the fluid starts flowing, $r \sim d/2$, $\Delta h=h$ and the threshold pressure, given by
the condition $d^2r/dt^2 \geq 0$, can be written
\begin{equation} \label{eq:dPY}
\delta P_Y = \frac{4\gamma}{d} + \rho g h + \frac{2G'}{d} \left( \frac{d}{2}-r_n \right) .
\end{equation} 
Before the emission of the first bubble in the cascade, $r_0 = 0$ and we can estimate $\delta P_Y \sim 180$~Pa, which is
consistent with the pressure signal measured in the experiments (Fig.~\ref{fig:cascade}, top).

\begin{figure}[t]
\begin{center}
\includegraphics[width=0.9\columnwidth]{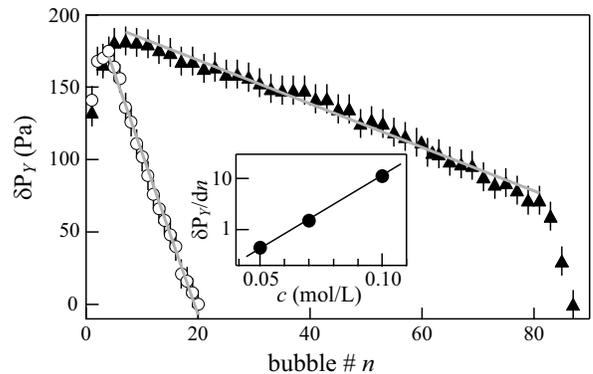}
\end{center}
\caption{Evolution of $\delta P_y$ as a function of the bubble number in the cascade.
In the main part of the cascade, $\delta P_y$ is a linear decaying function of the bubble
number $n$ (Symbol, $c$~[mol.L$^{-1}$]): ($\circ,0.1$), ($\blacktriangle,0.07$) 
[$V=147$~mL, $Q=0.65$~mL/s].}
\label{fig:dPY}
\end{figure}

As the bubble properties vary continuously inside the cascade, we investigate the
evolution of $\delta P_Y$ as a function of the bubble number. We find that, in the 
main part of the cascade, $\delta P_Y$ is a linear, decaying function of 
the bubble number (Fig.~\ref{fig:dPY}). The corresponding slope, constant from
one cascade to the other, increases with the fluid concentration (Fig.~\ref{fig:dPY}, inset).
This linear relationship can be explained by a simple heuristic model, based on the key
ingredients of the Maxwell model (Fig.~\ref{fig:Maxwell}). 

We consider successive bubbles rising through or piercing the thin fluid layer. When the 
bubble rises during the characteristic time $t_b$, $r \sim d/2$ (Fig.~\ref{fig:Maxwell}) 
and we can write the differential equation describing the temporal evolution of $r_n$:
\begin{equation}
G' \left( \frac{d}{2}- r_n \right) = \eta \, \dot r_n
\end{equation} 
which gives, right after the bubble rise 
\begin{equation}
r_{n}(t_b)=\left( r_{n-1} - \frac{d}{2} \right) e^{-t_b/\tau} + \frac{d}{2}
\end{equation} 
where $r_{n-1}$ is the initial condition from which $r_n$ evolves and
$\tau = \eta / G'$ the characteristic time associated with the Maxwell fluid.

After the bubble rise, we suppose that the fluid layer closes almost immediately, due in particular to 
the surface tension. The fluid is hence at rest, and $r_n$ relaxes towards $0$ during a time 
$T-t_b$, where $T$ is the average time between bubble emission. For sake of convenience,
we consider here that $T$ and $t_b$ are the same for each bubble. 
During this stage of relaxation, $r_n$ obeys the following equation:
\begin{equation}
\tau \dot r_n +  r_n = 0 \, ,
\end{equation} 
which gives, after a time $t=T-t_b$, $r_n=r_{n}(t_b) e^{-t/\tau}$. We can therefore
write the recurrence equation giving the displacement after the $n^{th}$ bubble:
\begin{equation}
r_n = r_{n-1} e^{-T/\tau} + \frac{d}{2} e^{-T/\tau} \left( e^{-t_b/\tau} - 1 \right).
\end{equation} 
By recurrence, and considering that $r_0=0$, we thus get the equation describing the displacement $r_n$:
\begin{equation}
r_n =  \frac{d}{2} \left( e^{-t_b/\tau} - 1 \right) \left( \frac{e^{-T/\tau}-e^{-nT/\tau}}{1-e^{-T/\tau}} \right) .
\end{equation} 
For $t_b < \tau$ and $T < \tau$ (both the characteristic times of bubble rising through the fluid layer 
and bubble formation and emission are smaller than the Maxwell time), 
the slope $dr_n/dn$ can be approximated to $(d/2)(t_b/\tau)$. 
From Eq.~\ref{eq:dPY}, we thus obtain the following expression for the threshold pressure:
\begin{equation}
\delta P_Y \sim \left( \frac{4\gamma}{d} + \rho g h +G' \right) - \frac{1}{2} G' \left( \frac{t_b}{\tau}  \right) n
\end{equation} 
The average time for a bubble emission (formation and growth, regions 2 and 3 in Fig.~\ref{fig:cascade}, bottom) 
inside a cascade, for $c=0.1$~mol.L$^{-1}$, can be estimated to $t_b \sim 0.2$~s. The characteristic time $\tau=\eta/G'$  
is obtained from rheological measurements (see Appendix) and can be estimated to $\tau \sim 0.5$~s, with 
$\eta \sim 25$~Pa.s and $G' \sim 50$~Pa. We thus estimate a slope $-(G'/2) (t_b / \tau)$ of about $-10$, 
in agreement with the experimental results (see Fig.~\ref{fig:dPY} for $c=0.1$~mol.L$^{-1}$).

\subsection{Fluid deformation}

Above the threshold overpressure $\delta P_Y$, the fluid starts flowing and a bubble 
grows at the tip of the injection hole (Fig.~\ref{fig:cascade}, bottom, region~2). 
The overpressure then departs from its linear increase, as a consequence of the volume increase due to the bubble growth.

By deriving the general equation for an ideal gas, and denoting $v_b$ the volume of the bubble hence
formed, we get 
\begin{equation}
\frac{dP}{dt}+ \left( \frac{P}{V} \right) \frac{dv_b}{dt} = \left( \frac{RT}{V} \right) \frac{Q}{V_{mol}}
\end{equation}
Further integration leads to the general expression of the pressure variation in time:
\begin{equation}
P(t)= \left( \frac{RT}{V} \right) \left( \frac{Q}{V_{mol}} \right) t - \left( \frac{P}{V} \right) v_b
\end{equation}
where $R=8.314$~J~K$^{-1}$~mol$^{-1}$ is the ideal gas constant and $V_{mol}=24.7$~L
is the molar volume of air at 25$^\circ$C. 
Here we do not develop further the calculation, but note that the departure from the linear trend is 
linear with the bubble volume, which provides a rough estimation of this latter,
from about $v_b \sim 1$~mL to 10~mL from the beginning to the end of the cascade.

\subsection{Bubble emission}
\label{sec:emission}

Finally, the bubble reaches the free surface -- we remind here that the fluid layer is of the order
of the bubble size --, the gas is suddenly released and the overpressure quickly 
drops (Fig.~\ref{fig:cascade}, bottom, region~3). In order to get an estimate of
the characteristic time over which the pressure drops, we write, on the one hand, Bernoulli's equation
to describe the air flowing from the chamber of volume $V$ through the opening
of diameter $d$, up to the surface of the fluid layer: $1/2 \rho v^2=\delta P$, which
gives the flow velocity through the opening \cite{Batchelor00}:
\begin{equation} \label{eq:v}
v = \sqrt{\frac{2 \delta P}{\rho}} \, .
\end{equation}
On the other hand, by considering the air as an ideal gas, we get $\delta P/P \sim v_b/V$, 
where $V$ is the initial gas volume, equal to the chamber volume, and $v_b$ the volume variation
corresponding to the volume of the bubble connected to the injection point.
The typical time to empty the chamber can be written $\tau^*=\delta V/Q_v$, where 
$Q_v=\pi (d^2/4)v$ is the volumetric flow-rate through the hole.
By using Eq.~\ref{eq:v}, we get the characteristic time for the pressure drop:
\begin{equation} 
\tau^*= \left( \frac{V}{P} \right) \frac{1}{\pi d^2} \sqrt{\frac{\rho_a}{32 \delta P}}
\end{equation}
where $\rho_a=1.2$~kg.m$^{-3}$ is the air density, and $P=10^5$~Pa the atmospheric
pressure. $\tau^*$ is of the order of a few milliseconds, compatible with the measurements
of the pressure drop (region~3 in Fig.~\ref{fig:cascade}, bottom).

Note that the drastic pressure decrease due to the hole opening is followed by oscillations 
observed right after the bubble emission, which correspond to the elastic response of the 
fluid to the sudden stress imposed by the closing of the bubble walls, 
after the air release \cite{Berret97}.

\subsection{Open channel lifetime}
\label{sec:channel}

Here, we develop qualitative arguments to estimate the lifetime of the open
channel. We note $r^*$ the hole radius for which the channel remains open at 
the end of a cascade, and $r^*_n$ the associated value of the inner displacement
from the Maxwell model developped above (Fig.~\ref{fig:Maxwell}). When the channel from the injection
nozzle to the fluid free surface remains open, $d^2r/dt^2=0$ and we can write, 
based on Eq.~\ref{eq:motion}:
\begin{equation}  \label{eq:aperture}
-\frac{4\gamma}{d} - G' \left( \frac{d/2-r^*_n}{d} \right) + \delta P - \rho g h = 0
\end{equation}
and
\begin{equation}  \label{eq:rstar}
-\frac{2\gamma}{r^*} - G' \left( \frac{r^*-r^*_n}{d} \right) - \rho g h = 0
\end{equation}
Eq.~\ref{eq:aperture} provides the value of $r^*_n$ after the last bubble in the cascade,
while Eq.~\ref{eq:rstar} gives the channel radius right after the pressure drop in the system.
By substracting Eqs.~\ref{eq:aperture} and \ref{eq:rstar} and neglecting the capillary forces 
when the channel if fully open,
we find that the channel remains open when the overpressure $\delta P \sim G'$.
This rough approximation is consistent with the experimental value. Indeed, from
Fig.~\ref{fig:cascade}, the channel remains open once the overpressure
reaches a value close to 50~Pa, of the order of $G'$ for an excitation frequency
of the order of the frequency of bubbles rising through the fluid ($\omega \sim 1$~Hz, 
see Appendix~\ref{app:modulii}).

After the channel opening, we can write the balance between the main forces at stake,
the weight and viscous dissipation, $\eta \dot r /d \sim \rho g h/2$. 
The characteristic lifetime of the open channel can therefore be roughly estimated 
to $\tau_c \sim 2\eta / \rho g h$, of the order of a second.
Experimentally, $\tau_c$ is of a few seconds. This discrepancy can be explained 
first, by the rough approximation for the radius dynamic equation; then, by the
effective fluid viscosity which, for micelles, can be larger during elongational flowing 
than the viscosity measured under shear \cite{Prudhomme94,Walker96}.

\section{From cascades to rheology}
\label{sec:rheology}

\begin{figure}[t]
\begin{center}
\includegraphics[width=0.9\columnwidth]{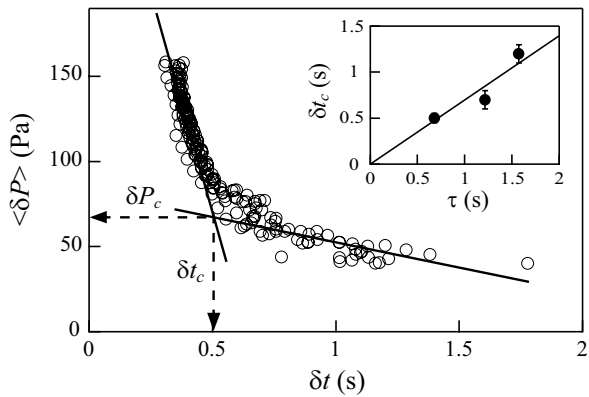}
\end{center}
\caption{Mean pressure drop for a bubble vs. time interval for bubble emission.
[$V$=147~mL, $h$=5~mm, $c$=0.1~mol.L$^{-1}$, $Q$=0.65~mL/s].
{\it Inset:} Variations of $\delta t_c$ with the fluid characteristic time $\tau$, determined
here as $\tau=G'/G"\omega$ (see text). 
From left to right: $c=0.1, 0.07$ and $0.05$~mol.L$^{-1}$.}
\label{fig:dPvsdt}
\end{figure}

The bubble emission characteristics vary continuously inside the cascade: through time, 
a bubble inside the cascade is emitted with a smaller pressure drop $\delta P$, and 
a longer emission time $\delta t$ (Fig.~\ref{fig:cascade}). Note that for the heuristic model
developped in section~\ref{sec:dPY}, we considered $\delta t $ roughly constant 
and equal to $T=\langle \delta t\rangle$. In this section, we investigate the variations
of both $\delta t$ and $\delta P$ for the successive bubbles inside one cascade.

Figure~\ref{fig:dPvsdt} displays the mean value of the overpressure $\langle \delta P \rangle$, 
over a bubble release, as a function of the time interval $\delta t$ over which the same 
bubble is emitted, for a series of bubble cascades.
All data from different cascades, from the same experimental series, have been 
superimposed. Over different cascades, the data all collapse in the
same curve, for a given set of parameters ($c$, $h$, $Q$, $V$).
The curves remain unchanged when both the volume $V$ and the flow rate $Q$ vary. 

Two different regimes can be distinguished, at short and long time scales, respectively
(Fig.~\ref{fig:dPvsdt}, solid lines).
The limit between both regimes defines a pressure and time threshold, respectively
$\delta P_c$ and $\delta t_c$. 
A statistical study over different volumes ($ 56 \leq V \leq 161$~mL) and flow rates 
($0.2 \leq Q \leq 1.2$~mL/s) gives a constant pressure threshold $\delta P_c = 71.6 \pm 5.2$~Pa.
By comparing the characteristic time $\delta t_c$ to the Maxwell characteristic time $\tau$ of the fluid 
(see Appendix), we get the direct relationship 
\begin{equation}
\delta t_c \sim \tau \, . 
\end{equation}
This relationship holds true for the three different fluid concentrations where bubble cascades
are observed (Fig.~\ref{fig:dPvsdt}, inset).
Measuring the characteristic time linked with the bubble cascades therefore provides 
a direct signature on the fluid rheology.

\section{Conclusion}
\label{sec:conclusion}

Injecting air through a thin layer of micellar fluid displays a wide range of dynamic behavior.
We report the existence of a peculiar degassing regime, the {\it bubble cascades} regime, 
for which the air is released via successive bubbles which properties (maximum overpressure 
and emission duration) vary continuously through time. This regime is observed over a
wide range of parameters (air flow-rate $Q$, chamber volume $V$ and fluid height $h$).
The cascades repeat periodically in time, separated by a few seconds during which 
a channel remains open between the injection nozzle and the fluid free surface, through 
which the air degass continuously. Measuring the overpressure at the injection point
makes it possible to investigate the different stages: pressure increase in the chamber, 
bubble formation and fluid flow, bubble emission and pressure drop. We find that 
the cascade periodicity and maximum overpressure depend neither on $Q$
nor $V$. The number of bubbles emitted per cascade depends linearly on the
injection flow-rate $Q$. All the different steps of the overpressure evolution 
can be explained by a simple heuristic model, following the classical Maxwell
description of a viscoelastic fluid.

Finally, we underline the interest of such simple experiment. In addition to reporting a new
phenomenon (the bubble cascades), we point out that measuring the evolution of the 
overpressure inside the cascades provides a direct insight into the fluid viscoelastic characteristic time,
linked with its rheology. Further work will concentrate on the microscopic behavior
of the micelles. In particular, the time over which the rising bubble shears the fluid (typically,
0.2~s) suggests a partial alignment of the micelles (see Fig.~\ref{fig:rheology}, Appendix~\ref{app:rheology}). 
Visualizing the fluid by birefringence will make it possible to determine
if the micelles are aligned through time by the shear flow generated by the successive bubbles,
and how this alignment is linked with the local rheology. We also propose to investigate
the effective fluid viscosity under elongation, which can be much larger than the shear viscosity,
and could explain the discrepancy between the estimated and measured time for the open
channel lifetime (section~\ref{sec:channel}).

\appendix

\section{Flow curve}
\label{app:rheology}

The rheology of the CTAB/NaSal mixtures is characterized by measurements on 
two different rheometers: C-VOR 150, Bohlin Instruments and AR1000, TA Instruments. 
All measurements are performed with a plate-plate geometry. Sand paper is glued to the plates 
in order to prevent any sliding at the walls (typical rugosity of the order of 1~$\mu$m). 
The following results do not aim at a full rheological characterization
of the samples used in our experiments. They only provide the general mechanical
behavior of the different mixtures, and a support for the interpretation proposed
in section~\ref{sec:rheology}. More detailled informations on the rheology of such
systems can be found, for instance, in 
\cite{Shikata87,Shikata88,Shikata89,Cates90a,Cates90b,Berret94,Berret97,Inoue05,Lerouge10}.

\begin{figure}[t]
\begin{center}
\includegraphics[width=0.9\columnwidth]{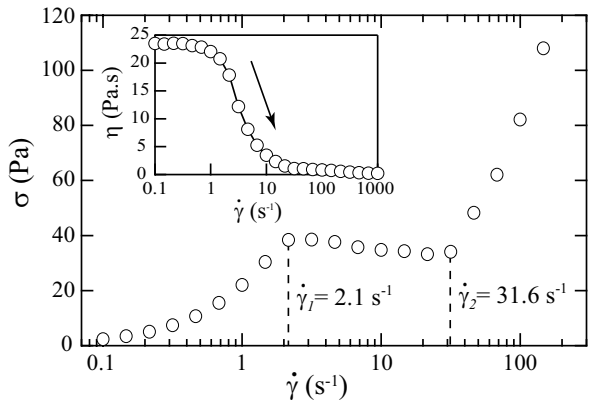}
\end{center}
\caption{The flow curve, shear stress $\sigma$ vs shear rate $\dot \gamma $ obtained by increasing 
$\dot \gamma$ [$c=0.1$~mol.L$^{-1}$].
{\it Inset:} Viscosity as a function of shear rate. The arrow indicates the shear-thinning behavior
[C-VOR 150 Bohlin rheometer, plate-plate geometry, diameter 60~mm, gap 400~$\mu$m, 
waiting time $60$~s per point].}
\label{fig:rheology}
\end{figure}

The semi-dilute solution behaves as a Newtonian fluid at low shear rate 
($\dot \gamma < \dot \gamma_1 = 2.1$~s$^{-1}$), with a viscosity plateau
$\eta \sim 25$~Pa, and exhibits non-Newtonian
properties for higher shear rate (Fig.~\ref{fig:rheology}). The flow curve is classical for
micellar fluids under shear \cite{Berret97,Lerouge10}, with a plateau in $\sigma$ vs. 
$\dot \gamma$ between $\dot{\gamma_1}=2.1$~s$^{-1}$ and 
$\dot{\gamma_2}=31.6$~s$^{-1}$. The first transition in the flow curve ($\dot \gamma_1$) 
provides an access to the characteristic time associated with the viscoelastic Maxwell model (Fig.~\ref{fig:Maxwell}),
$\tau = 1 / \dot \gamma_1 \sim 0.5$~s.

For $\gamma<\dot{\gamma_1}$, the flow is homogeneous and isotropic; 
for $\dot{\gamma_1}<\gamma<\dot{\gamma_2}$, shear-bands appear and the flow is 
strongly inhomogeneous; for $\gamma>\dot{\gamma_2}$, finally, the flow is homogeneous and 
nematic, with the micelles aligned in the shear direction \cite{Berret97}. As the typical time
for a bubble to rise up the fluid layer is $t_b \sim 0.2$~s (section~\ref{sec:dPY}), 
we expect to be in the stress plateau and, therefore, to observe a partial alignment 
of the micelles (section~\ref{sec:conclusion}).

\section{Elastic and viscous modulii}
\label{app:modulii}

\begin{figure}[t]
\begin{center}
\includegraphics[width=0.9\columnwidth]{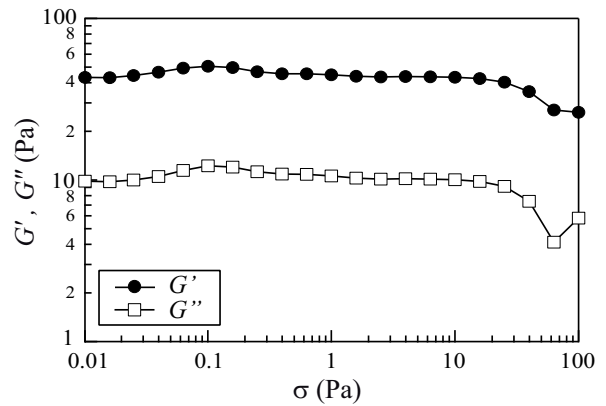}
\end{center}
\caption{Elastic ($G'$) and viscous ($G''$) modulii as a function of the applied stress for
the CTAB/NaSal mixture at $c=0.1$~mol.L$^{-1}$
[oscillation test $\omega=1$~Hz, AR1000 rheometer, plate-plate geometry].}
\label{fig:moduliiSigma}
\end{figure}

\begin{figure}[t]
\begin{center}
\includegraphics[width=0.9\columnwidth]{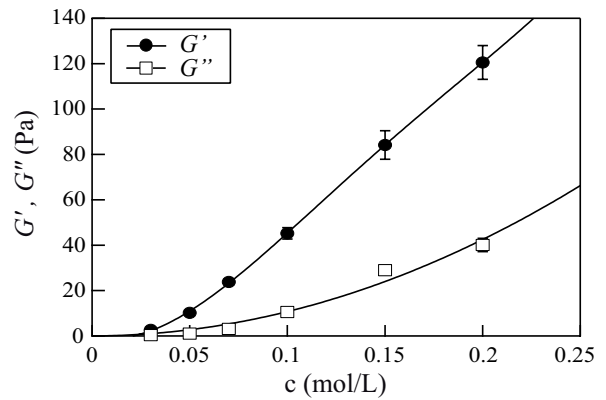}
\end{center}
\caption{Average elastic ($G'$) and viscous ($G''$) modulii corresponding to the plateau value in 
Fig.~\ref{fig:moduliiSigma}] as a function of the fluid concentration. Lines are given as eyeleads.}
\label{fig:moduliiC}
\end{figure}

Figure~\ref{fig:moduliiSigma} displays the elastic ($G'$) and viscous ($G"$) moduli for the 
fluid $c=0.1$~mol.L$^{-1}$, for an oscillation test ($\omega=1$~Hz, typically the period between
two bubbles in our experiments). We observe that over a large range of applied stress,
the elastic and viscous modulii are constant. 
The average plateau values of  $G'$ and $G"$ displayed in Fig.~\ref{fig:moduliiSigma} provide 
another estimation of the Maxwell time:
\begin{equation}   \label{eq:tauG}
\tau=\frac{G'}{G'' \omega}
\end{equation}
We find $\tau \sim 0.5$~s for $c=0.1$~mol.L$^{-1}$, consistent with the
estimation from the previous viscosity measurements (Appendix~\ref{app:rheology}).

The average plateau value strongly increases
as a function of the fluid concentration (Fig.~\ref{fig:moduliiC}). 
The values of $\tau$ for different fluid concentrations (Fig.~\ref{fig:dPvsdt} inset, 
section~\ref{sec:dPY}) have been estimated from the oscillatory measurements and 
Eq.~\ref{eq:tauG}.

Note, finally, that another estimation of the characteristic time $\tau$ can be obtained by
\begin{equation}
\tau=\frac{\eta}{G'}
\end{equation}
which, again, gives $\tau \sim 0.5$~s, in agreement with the above estimations.

 

\begin{thebibliography}{0}%
\makeatletter
\providecommand \@ifxundefined [1]{%
 \@ifx{#1\undefined}
}%
\providecommand \@ifnum [1]{%
 \ifnum #1\expandafter \@firstoftwo
 \else \expandafter \@secondoftwo
 \fi
}%
\providecommand \@ifx [1]{%
 \ifx #1\expandafter \@firstoftwo
 \else \expandafter \@secondoftwo
 \fi
}%
\providecommand \natexlab [1]{#1}%
\providecommand \enquote  [1]{``#1''}%
\providecommand \bibnamefont  [1]{#1}%
\providecommand \bibfnamefont [1]{#1}%
\providecommand \citenamefont [1]{#1}%
\providecommand \href@noop [0]{\@secondoftwo}%
\providecommand \href [0]{\begingroup \@sanitize@url \@href}%
\providecommand \@href[1]{\@@startlink{#1}\@@href}%
\providecommand \@@href[1]{\endgroup#1\@@endlink}%
\providecommand \@sanitize@url [0]{\catcode `\\12\catcode `\$12\catcode
  `\&12\catcode `\#12\catcode `\^12\catcode `\_12\catcode `\%12\relax}%
\providecommand \@@startlink[1]{}%
\providecommand \@@endlink[0]{}%
\providecommand \url  [0]{\begingroup\@sanitize@url \@url }%
\providecommand \@url [1]{\endgroup\@href {#1}{\urlprefix }}%
\providecommand \urlprefix  [0]{URL }%
\providecommand \Eprint [0]{\href }%
\@ifxundefined \urlstyle {%
  \providecommand \doi  [0]{\begingroup \@sanitize@url \@doi}%
  \providecommand \@doi [1]{\endgroup \@@startlink {\doibase
  #1}doi:\discretionary {}{}{}#1\@@endlink }%
}{%
  \providecommand \doi  [0]{doi:\discretionary{}{}{}\begingroup
  \urlstyle{rm}\Url }%
}%
\providecommand \doibase [0]{http://dx.doi.org/}%
\providecommand \Doi [0]{\begingroup \@sanitize@url \@Doi }%
\providecommand \@Doi  [1]{\endgroup\@@startlink{\doibase#1}\@@Doi}%
\providecommand \@@Doi [1]{#1\@@endlink}%
\providecommand \selectlanguage [0]{\@gobble}%
\providecommand \bibinfo  [0]{\@secondoftwo}%
\providecommand \bibfield  [0]{\@secondoftwo}%
\providecommand \translation [1]{[#1]}%
\providecommand \BibitemOpen [0]{}%
\providecommand \bibitemStop [0]{}%
\providecommand \bibitemNoStop [0]{.\EOS\space}%
\providecommand \EOS [0]{\spacefactor3000\relax}%
\providecommand \BibitemShut  [1]{\csname bibitem#1\endcsname}%
\end{thebibliography}%


\begin{thebibliography}{}

\bibitem{Gladden07}
J. Gladden and A. Belmonte, Phys. Rev. Lett. {\bf 98}, 224501 (2007).

\bibitem{Tabuteau09}
H. Tabuteau, S. Mora, G. Porte, M. Abkarian, and C. Ligoure, Phys. Rev. Lett. {\bf 102}, 155501 (2009).

\bibitem{Bird87}
R. B. Bird, R. C. Armstrong, and O. Hassager, {\it Dynamics of Polymeric Liquids} (Wiley, New York, 1987),
Vol. I and II .

\bibitem{Mewis79}
J. Mewis, J. Non-Newtonian Fluid Mech. {\bf 6}, 1 (1979).

\bibitem{Barnes97}
H. A. Barnes, J. Non-Newtonian Fluid Mech. {\bf 70}, 1 (1997).

\bibitem{Mewis09}
J. Mewis, and N.~J. Wagner, Adv. Coll. Inter. Sci. {\bf 147-148}, 214 (2009) .

\bibitem{Coussot02}
P. Coussot, Q. D. Nguyen, H. T. Huynh, and D. Bonn, Phys. Rev. Lett. {\bf 88}, 175501 (2002).

\bibitem{Moller06}
P. C. F. M{\"o}ller, J. Mewis, and D. Bonn, Soft Matter {\bf 2}, 274 (2006).

\bibitem{Shikata88}
T. Shikata, H. Hirata, and T. Kotaka, Langmuir {\bf 4}, 354 (1988).

\bibitem{Cates90b}
M. Cates and S. Candau, J. Phys.: Condens. Matter {\bf 2}, 6869 (1990).

\bibitem{Gelbart94}
W.M. Gelbart, {\it Micelles, Membranes, Microemulsions and Monolayers}
(Springer, New York, 1994).

\bibitem{Cates90a}
M. Cates, J. Phys. Chem. {\bf 94}, 371 (1990).

\bibitem{Berret97}
J. F. Berret, Langmuir {\bf 13}, 2227 (1997).

\bibitem{Chhabra07}
R. P. Chhabra, {\it Bubbles, Drops and Particles in non-Newtonian Fluids}
(Chemical industries series \#113, Taylor \& Francis, 2007), 2nd edition.

\bibitem{Belmonte00}
A. Belmonte, Rheol. Acta {\bf 39}, 554 (2000).

\bibitem{Handzy04}
N. Handzy and A. Belmonte, Phys. Rev. Lett. {\bf 92}, 124501 (2004).

\bibitem{Divoux08b}
T. Divoux, V. Vidal, F. Melo, and J.-C. G\'eminard, Phys. Rev. E {\bf 77}, 006310 (2008).

\bibitem{Jayaraman03}
A. Jayaraman, and A. Belmonte, Phys. Rev. E {\bf 67}, 065301(R) (2003).

\bibitem{Hassager79}
O. Hassager, Nature {\bf 279}, 402 (1979).

\bibitem{Daugan02a}
S. Daugan, L. Talini, B. Herzhaft, and C. Allain, Eur. Phys. J. E {\bf 7}, 73 (2002).

\bibitem{Daugan02b}
S. Daugan, L. Talini, B. Herzhaft, and C. Allain, Eur. Phys. J. E {\bf 9}, 55 (2002).

\bibitem{Li01}
H. Z. Li, X. Frank, D. Funfschilling, and Y. Mouline, Chem. Eng. Sci. {\bf 56}, 6419 (2001).

\bibitem{Lin03}
T.-J. Lin and G.-M. Lin, Can. J. Chem. Eng. {\bf 81}, 476 (2003).

\bibitem{Li04b}
H. Z. Li, X. Frank, D. Funfschilling, and P. Diard, Phys. Lett. A {\bf 325}, 43 (2004).

\bibitem{Divoux09}
T. Divoux, E. Bertin, V. Vidal, and J.-C G\'eminard, Phys. Rev. E {\bf 79}, 056204 (2009).

\bibitem{Gostiaux02}
L. Gostiaux, H. Gayvallet, and J.-C G\'eminard, Gran. Matt. {\bf 4}, 39 (2002).

\bibitem{Varas09}
G. Varas, V. Vidal, and J.-C. G\'eminard, Phys. Rev. E {\bf 79}, 021301 (2009)

\bibitem{Kliakhandler02}
I.~L. Kliakhandler, Phys. Fluids {\bf 14}, 3375 (2002).

\bibitem{Shikata87}
T. Shikata, H. Hirata, and T. Hotaka, Langmuir {\bf 3}, 1081 (1987).

\bibitem{Inoue05}
T. Inoue, Y. Inoue, and H.Watanabe, Langmuir {\bf 21}, 1201 (2005).

\bibitem{RMQ}
When the fluid concentration increases, the characteristic time associated with the Maxwell model increases 
(see Appendix). For the same excitation (bubble rising), the fluid behavior gets closer to
an elastic solid response at short time (easier channel opening); at long time, due to the higher viscosity, 
it is more difficult to close the channel, which is therefore more easily sustained.

\bibitem{Batchelor00}
G.~K. {Batchelor}, An Introduction to Fluid Dynamics, {\it Cambridge Univ. Press} (2000).

\bibitem{Prudhomme94}
R.~K. {Prud'homme}, and G.~G. Warr, Langmuir {\bf 10}, 3419 (1994).

\bibitem{Walker96}
L.~M. {Walker}, and P. Moldenaers, Langmuir {\bf 12}, 6309 (1996).

\bibitem{Shikata89}
T. Shikata, H. Hirata, and T. Kotaka, Langmuir {\bf 5}, 398 (1989).

\bibitem{Lerouge10}
S. Lerouge and J.-F. Berret, in {\it Polymer Characterization: Rheology, Laser Interferometry, 
Electrooptics}, edited by K. Dusek and J.-F. Joanny (Springer-Verlag, 2010) pp. 1--171.

\bibitem{Berret94}
J. F. Berret, D. C. Roux, and G. Porte, J. Phys. II France {\bf 4}, 1261 (1994).


\end{thebibliography}

\end{document}